\documentclass[prl,aps,twocolumn]{revtex4}  
\usepackage{epsf}
\makeatletter  
  
\def\compoundrel#1\over#2{\mathpalette\compoundreL{{#1}\over{#2}}}  
\def\compoundreL#1#2{\compoundREL#1#2}  
\def\compoundREL#1#2\over#3{\mathrel  
  {\vcenter{\hbox{$\m@th\buildrel{#1#2}\over{#1#3}$}}}}  
\makeatother  
  
\begin{document}  
  
%\draft  
  
\title{A quantum gate array can be programmed to evaluate  
the expectation value of any operator}

\author{Juan Pablo Paz$^{1,2}$ and Augusto Roncaglia$^1$}
  
\affiliation{(1) Departamento de F\'{\i}sica ``J.J. Giambiagi'',  
FCEN, UBA, Pabell\'on 1, Ciudad Universitaria, 1428 Buenos Aires, Argentina}  
\affiliation{(2) Theoretical Division, LANL, MSB213, Los Alamos, NM 87545, USA}
  
\begin{abstract}  
A programmable gate array is a circuit whose action is controlled 
by input data. %It has been shown that a general purpose programmable 
%quantum computer does not exist. However in this letter 
In this letter we describe a special--purpose quantum circuit that can be 
programmed to evaluate 
the expectation value of any operator $O$ acting on a space of states 
of $N$ dimensions. 
The circuit has a program register whose state $|\Psi(O)\rangle_P$ 
encodes the operator $O$ whose expectation value is to be evaluated. 
The method requires knowledge of the expansion of $O$ in a basis of the
space of operators. We discuss some applications of this circuit and its  
relation to known instances of quantum state tomography.
\end{abstract}

\date{\today}  
  
%%%%%%%%%%%%%%%%%%%%%%%%%%%%%%%%%%%%%%%%%%%%%%%%%%%%%%%%%%%  
%  
%		Main text  
%  
%%%%%%%%%%%%%%%%%%%%%%%%%%%%%%%%%%%%%%%%%%%%%%%%%%%%%%%%%%%  

\maketitle
%\narrowtext  

An important feature of classical computers is that they can be programmed.
That is to say, a fixed universal device can perform different tasks
depending on the state of some input registers. These registers 
define the program the device is executing. Quantum computers 
\cite{NielsenChuang} have a rather different property: Thus, Nielsen and 
Chuang established in \cite{NielsenChuangPRL} that a general purpose 
programmable quantum computer does not exist. Such a device would 
have to have the following features: It should consist of a 
fixed gate array with a data register and a program 
register. The array should work in such a way that the state of the 
program register encodes the unitary operator ${\cal U}$ that is 
applied to the state of the data register. As shown in  
\cite{NielsenChuangPRL}, such devices cannot be universal since 
different unitary operators require orthogonal states of the program
register. However, some interesting examples of programmable devices
could still be constructed. For example, non-deterministic programmable 
gate arrays were first considered in \cite{NielsenChuangPRL} 
and analyzed later in a variety of examples 
\cite{ProgramableArrays}. More recently, quantum ``multimeters'' 
were introduced and discussed  in \cite{DusekBuzek}. 
Such devices are fixed gate arrays acting on a data register and 
a program register, together with a final fixed 
projective measurement on the composite system. They are programmable 
quantum measurement devices \cite{DusekBuzek}
that act either non--deterministically or in an approximate way 
(see \cite{FiurasekDusekFilip}).

In this letter we will describe a different kind of programmable quantum gate
array that is useful to solve the following problem: 
Suppose that we are given an operator $O$ acting on an $N$ dimensional 
Hilbert space and a quantum state $\rho$. By this we mean that someone 
supplies us with many copies of a quantum system prepared in the same 
state $\rho$ and defines for us the operator $O$ by specifying its expansion 
in a basis of the space of operators. Our task is 
to compute the expectation value of $O$ in the state $\rho$. 
We will show that it is possible to construct a programmable
circuit that evaluates such expectation value by measuring the polarization
of a single qubit. The inputs of such circuit are 
a data register, a program register and an auxiliary qubit. The circuit 
evaluates the expectation value of an operator $O$ (specified by the 
program register) in the quantum state $\rho$ of the data register. 
The expectation value ${\rm Tr}(\rho O)$ is obtained by performing a 
measurement of the polarization of the auxiliary qubit. We will 
describe how to construct these circuits and exhibit an 
interesting example: a programmable array to efficiently solve a 
class of quantum decision problems concerning properties of quantum 
phase space distributions. 

The quantum gate arrays discussed in this letter are designed 
using the ``scattering circuit'' shown in Figure 1 as a simple primitive. 
In such circuit a  system, initially in the state $\rho$, is brought 
in contact with an ancillary qubit prepared in the state $|0\rangle$. 
This ancilla acts as a probe particle in a scattering experiment. 
The algorithm consist of the following steps: i) Apply a Hadamard 
transform $H$ to the ancillary qubit. Since 
$H|0\rangle=(|0\rangle+|1\rangle)/\sqrt 2$,
$H|1\rangle=(|0\rangle-|1\rangle)/\sqrt 2$, the new state of the qubit
is $(|0\rangle+|1\rangle)/\sqrt 2$. ii) Apply a ``controlled--$A$''
operator, which does nothing if the state of the ancilla is
$|0\rangle$ and applies the unitary operator $A$ to the system if 
the ancilla is in state $|1\rangle$, iii) Apply another Hadamard 
gate to the ancilla and perform measurements of its spin 
polarizations along the $z$- and $y$-axes. Given sufficiently 
many independent instances of the experiment, the measurements yield 
the expectation values $\langle\sigma_z\rangle$ and 
$\langle\sigma_y\rangle$ of the Pauli spin operators $\sigma_z$ and
$\sigma_y$. This algorithm has the following remarkable property:
\begin{equation}
\langle\sigma_z\rangle=\mbox{Re}[ \, \mbox{Tr}(  A \rho) \, ], \
\langle\sigma_y\rangle=\mbox{Im}[ \, \mbox{Tr}(  A \rho) \, ].
\label{polarization}
\end{equation}
Different versions of this circuit play an important role in 
many quantum algorithms \cite{kitaev:qc1997a,cleve:qc1997b,Lloyd99,Knill98}. 
In particular, the scattering circuit was recently used as a basic 
tool to interpret tomography and spectroscopy as two dual forms of the 
same quantum computation \cite{NatureUS}. 

For our purpose it is useful to 
review how to use this scattering circuit as a primitive to design 
a tomographer (i.e., a device that after a number of experiments 
determines the quantum state of the system). As a consequence of 
equations (\ref{polarization}) we see that 
every time we run the algorithm for a known operator
$A$, we extract information about the state $\rho$. Doing so for a
complete basis of operators $\{A(\alpha)\}$ one gets complete
information and determines the full density matrix. Different
tomographic schemes are characterized by the basis of operators
$A(\alpha)$ they use. Of course, completely determining the quantum
state requires an exponential amount of resources. In fact, 
if the dimensionality of the Hilbert space of the system is $N$ 
then the complete determination of the quantum state involves running the 
scattering circuit for a complete basis of $N^2$ operators $A(\alpha)$. 
However, evaluating
any coefficient of the decomposition of $\rho$ in a given basis can be
done efficiently provided that the operators $A(\alpha)$ can be implemented
by efficient networks. A convenient basis set is defined as
(see, for example, \cite{NatureUS,MPS02}):
\begin{equation}
A(\alpha)= A(q,p)= U^q   R V^{-p}
\exp(i\pi pq/N).\label{aqp}
\end{equation}
Here, both $q$ and $p$ are integers between $0$ and $N-1$, 
$U$ is a cyclic shift operator in the computational basis
($U|n\rangle=|n+1\rangle$), $V$ is the cyclic shift operator 
in the basis related to the computational one via the discrete Fourier
transform, and $R$ is the reflection operator ($R|n\rangle=|N-n\rangle$). 
It is straightforward to show that the operators $A(q,p)$ are  
hermitian, unitary and form a complete orthonormal basis of the 
space of operators satisfying
\begin{equation}
\mbox{Tr}[A(\alpha) A(\alpha') ] = N \, \delta_N(q'-q) \delta_N(p'-p),
\end{equation}
where $\delta_N(x)$ is the periodic Kronecker delta function that is 
equal to one if $x=0$ (modulo $N$) and vanishes otherwise 
(the above operators form a ''quorum'', as defined in \cite{Dariano1}). 
With this choice for $A(q,p)$ 
the scattering circuit directly evaluates the discrete Wigner function 
\cite{NatureUS,Wooters,Leonhardt,MPS02}. 

\begin{figure}
 \centering \leavevmode
 \epsfxsize 3.2in
 \epsfbox{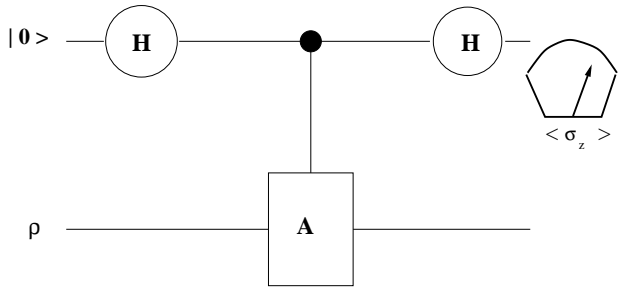}
\vspace{0.25 cm}
\caption{The scattering circuit that can be used to evaluate 
real and imaginary parts of the expectation value $Tr(\rho A)$ for 
a unitary operator $A$. $H$ denotes a Hadamard transform. The 
``controlled-A'' operation is such that 
$({\rm ctrl-}A)|q\rangle|\Psi\rangle=|q\rangle A^q|\Psi\rangle$.}
\label{scattering}
\end{figure}

We will now show how to design a programmable 
gate array to evaluate the expectation value of any operator $O$. 
We will assume that we know how to expand $O$ in a basis such as the one
used above: $O=\sum_{q,p} o(q,p) A(q,p)$. As the operators $A(q,p)$ are 
not only unitary but also hermitian, the real and imaginary parts 
of the complex coefficients $o(q,p)$ define the expansion of the 
hermitian and anti-hermitian pieces of $O$ in the basis $A(q,p)$. 
The expectation value of these two pieces can be evaluated separately
using the procedure described below (the results can then be 
combined to get the expectation of $O$). So, in what follows 
we will assume that the operator at hand is hermitian and that the 
coefficients of its expansion in the basis $A(q,p)$ are real numbers. 
To introduce our method, it is convenient to notice first that the 
evaluation of the expectation value of the operators $A(q,p)$ can be done 
using a programmable circuit that is independent of $q$ and $p$. 
Such circuit is illustrated in Figure 2. This is an application
of the scattering circuit shown in Figure 1 with two program 
registers used to encode the value of $q$ and $p$. When the quantum 
state of the program is $|\Psi\rangle_P=|q\rangle|p\rangle$ the circuit 
evaluates the expectation value of $A(q,p)$. This is accomplished by
letting the program registers to act as controls of the 
operators $U$ and $V$ (which, as mentioned, generate cyclic shifts
either in the computational or in the conjugated basis). 
Thus, the action of the circuit is such that 
when the auxiliary qubit is in state $|\sigma\rangle$ ($\sigma=0,1$)
and the state of the program is $|q\rangle|p\rangle$, the operator 
$A^\sigma(q,p)$ is applied to the system register. 
The network in Figure 2 is efficient 
since it can be built using a number of elementary gates which scales 
polynomially with $\log(N)$ \cite{MPS02}. 

The circuit has an obvious property: Different 
states  $|q\rangle|p\rangle$ are used to program the evaluation 
of the expectation value of orthogonal operators $A(q,p)$. 
It is clear that by restricting to such program states one has no real 
advantage with respect to the case in which $q$ and $p$ are stored as 
classical information. However, we can use more general program states: 
If the program register is in the state 
$|\Psi\rangle_{P}=\sum_{q,p}c(q,p) |q\rangle|p\rangle$ then
the same circuit evaluates the expectation value of a 
linear combination of the operators $A(q,p)$ since the final 
polarization is: 
\begin{equation}
\langle\sigma_z\rangle={\rm Tr}\left(\rho 
\sum_{q,p} |c(q,p)|^2 A(q,p) \right)
\label{program}
\end{equation}

Equation (\ref{program}) shows that this algorithm can be used to 
evaluate the expectation value of any operator that can be written as
a convex sum of the basis set $A(q,p)$. This is not general enough
since the expansion of a hermitian operator can include negative 
coefficients. For this purpose the method can be extended as follows: 
The most general hermitian operator can be expressed in the basis $A(q,p)$ 
as $O=\sum_{q,p} c^2(q,p) \exp(i\pi\phi(q,p))A(q,p)$, where 
$c(q,p)$ is a real number and $\phi(q,p)$ is either $0$ or $1$ 
($\phi(q,p)$ simply stores the information about the sign of each 
coefficient). We will assume that $\sum_{q,p}c^2(q,p)= 1$ (if this 
is not the case we can always renormalize the coefficients). 
For operators with some negative coefficients
we can use a third register consisting of a single qubit to store 
$\phi(q,p)$. The circuit evaluating the expectation value of $O$ 
is shown in Figure 3. The first two program registers store $q$ 
and $p$ and are used exactly in the same way as above. The third one, 
storing $\phi(q,p)$, is acted upon with a $\sigma_z$ operator, 
introducing the required phase $\exp(i\pi\phi(q,p))$. Then, if the 
state of the program register is
$|\Psi\rangle_P=\sum_{q,p}c(q,p)|q\rangle|p\rangle|\phi(q,p)\rangle$. 
the final polarization measurement turns out to be
\begin{equation}
\langle\sigma_z\rangle={\rm Tr}\left(\rho
\sum_{q,p} c^2(q,p) {\rm e}^{i\pi\phi(q,p)}A(q,p)\right)={\rm Tr}\left(\rho
O\right).
\label{program2}
\end{equation}

\begin{figure}
 \centering \leavevmode
 \epsfxsize 3.2in
 \epsfbox{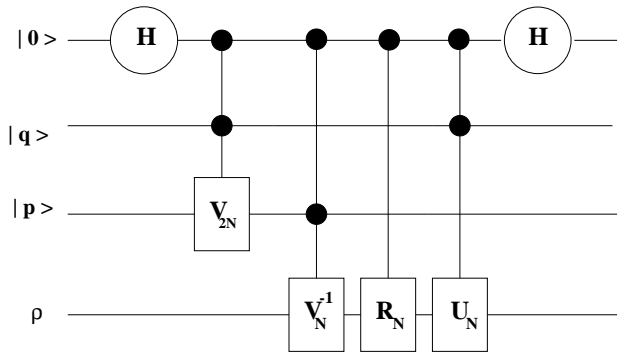}
\vspace{0.25 cm}
\caption{Programmable gate array evaluating ${\rm Tr}(\rho A(q,p))$ from
the polarization of the first qubit. 
The program state is $|\Psi\rangle_P=|q\rangle|p\rangle$. 
All ``controlled--$O$'' operators act as: 
$({\rm ctrl-}O)|n\rangle|\Psi\rangle=|n\rangle O^n|\Psi\rangle$.
$U_K$ ($V_K$) are cyclic shift operators in the computational (conjugated)
basis of a $K$ dimensional space. A subscript in an operator denotes
the dimensionality of the space in which it acts}
\label{circ2}
\end{figure}
%\vspace{0.5cm}

\begin{figure}
 \centering \leavevmode
 \epsfxsize 3.2in
 \epsfbox{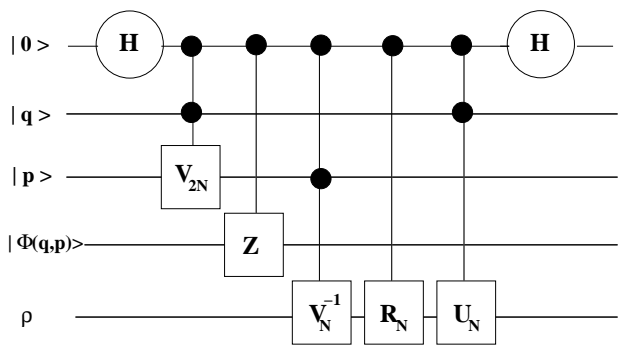}
\vspace{0.25 cm}
\caption{Programmable gate array evaluating ${\rm Tr}(\rho O)$ from the 
polarization of the first qubit. 
The program state is $|\Psi\rangle_P=\sum_{q,p}
c(q,p) |q\rangle |p\rangle |\Phi(q,p) \rangle$ where $c(q,p)$ and 
$\Phi(q,p)$ define the polar decomposition of the coefficients 
$o(q,p)={\rm Tr}(O A(q,p))/N$.}
\label{circ3}
\end{figure}

Summarizing, we showed that the measurement of the expectation value of 
any operator $O$ can be done using a programmable gate array. 
The hardware architecture is associated with the particular
choice of basis $A(q,p)$, which is just a matter of convenience, 
and is independent of $O$. The software used to program the 
array is obviously determined by the choice of hardware. The expectation
values of the hermitian and anti--hermitian parts of the operator $O$
are computed separately using a method that requires knowledge of 
the expansion of these operators in the basis $A(q,p)$. 
If the coefficients in the expansion are written as 
$o(q,p)= {\rm Tr}(O A(q,p))/N=c^2(q,p)\exp(i\pi\phi(q,p))$, 
the program state that needs to be prepared is 
$|\Psi\rangle_P=\sum_{q,p}c(q,p)\ |q\rangle|p\rangle|\phi(q,p)\rangle$ 
(where $\phi(q,p)=0$ or $1$). 
Thus, the coefficients of the expansion of $O$ in the 
basis $A(q,p)$ define the program state $|\Psi\rangle_P$ required 
to measure its expectation value. It is clear that in most cases 
the method will not be efficient. For example, both the task of defining the 
operator by specifying the coefficients $o(q,p)$ as well as the preparation 
of the program state $|\Psi\rangle_P$ are likely to be inefficient. 
The existence of efficient networks to implement ``controlled--$A(q,p)$'' 
operations is a less stringent conditions that is fulfilled by 
the basis defined in (\ref{aqp}) \cite{MPS02}. 
Having said this, it is worth noticing that there are 
sets of problems that can be efficiently solved using this 
method. We will now describe one such example. 

We will show that the circuit of Figure 2 can be easily adapted
to evaluate the sum of values of the Wigner function over various 
phase space domains. The program register is used to define the 
domain over which the Wigner function is averaged. If the domain 
is a line, the algorithm just evaluates the probability for the 
occurrences of the results of the measurement of a family of 
observables (see below). However, for more general domains (such as line
segments, parallelograms, etc) the circuit evaluates properties that
characterize a quantum state that cannot be simply casted 
in terms of probabilities. In this sense the circuit is as a programmable
tomographer measuring various features of phase space distributions. 
Before going into more details let us briefly review some properties 
of Wigner functions. 
Discrete Wigner functions can be used to represent the quantum 
state of a system in phase space \cite{NatureUS,Wooters,Leonhardt,MPS02}.  
For a system with an
$N$ dimensional space of states such function is defined
on a lattice of $2N\times 2N$ points $(q,p)$ where both $q$ and $p$ 
take values between $0$ and $2N-1$ (only $N\times N$ of these values 
are independent). At each phase space point the 
Wigner function is defined in terms of the operators $A(q,p)$ given 
in (\ref{aqp}) as 
\begin{equation}
W(q,p)={1\over 2N} {\rm Tr}(A(q,p)\rho). \label{wignerdef}
\end{equation}
As mentioned above, the measurement of this function can be done by 
using the scattering circuit \cite{NatureUS} or the programmable 
gate array of Figure 2. The program register encodes the value of 
$q$ and $p$. In general, if the program state is 
$|\Psi\rangle_P=\sum_{q,p}c(q,p)|q\rangle|p\rangle$, the final polarization 
measurement is $\langle\sigma_z\rangle=2N \sum_{qp} |c(q,p)|^2 W(q,p)$.
Thus, the program defines the region over which we 
average the value of the Wigner function. In general, preparing the 
program state associated with a general phase space region can 
be complicated. However, there are simple procedures to prepare the 
program states corresponding to general lines, segments and parallelograms. 
Let us begin with the simplest case: For the program state 
$|\Psi\rangle_P=\sum_q|q\rangle|p_0\rangle/\sqrt{2N}$, it is clear that 
the final polarization measurement reveals the sum of values of the 
Wigner function along the horizontal line defined as $p=p_0$: i.e., 
$\langle\sigma_z\rangle=\sum_q W(q,p_0)$. To consider more general lines
one can  notice that the horizontal line 
$p=c$ can be mapped into one satisfying the equation $p-bq=c$ (mod $2N$, 
with $b$ and $c$ integers between $0$ and $2N$) by applying a linear 
area preserving map. Unitary operators corresponding to quantizations 
of such maps (known as ''quantum cat maps'') act classically in phase
space. Thus, Wigner function flows from one point to another according
to the classical transformation. Using this feature, one can compute the 
sum of the Wigner function along any tilted line as follows: One 
first transforms the state with the appropriate cat map and later 
evaluates the sum of the Wigner function along a vertical or horizontal 
line. Moreover, the method can be made fully programmable by adding
extra registers to store the integers $b$ and $c$ parametrizing any 
cat map. 
Evaluating sums of Wigner functions over phase space lines is particularly 
interesting because of a crucial property of such functions: Thus, 
adding $W(q,p)$ along the line $ap-bq=c$ one obtains the 
probability to detect the eigenstate of the translation operator
$T(b,a)=U^aV^b\exp(i\pi ab/N)$ with eigenvalue $\exp(i\pi c/N)$. 
As a consequence, in this case the programmable gate array evaluates 
probabilities for the possible results of a set of measurements. 
Lines are a special case since more general phase space domains cannot 
be associated with projection operators. However, it is clear that 
the method described above can be applied to efficiently prepare 
program states for other phase space domains such as general (tilted) 
parallelograms: One first trivially prepare the program state for a 
parallelogram limited by vertical and horizontal segments and later 
tilt it applying the strategy based on the use of cat maps. Other 
simple phase space regions can also be programmed using variations of 
this method. It is also interesting to notice that using variations
of the circuit shown in Figure 3 we can also subtract values of the 
Wigner function in different phase space regions (which could be useful
if one is interested in comparing their values). 

In this letter we established the existence of a gate array that 
can be programmed to evaluate the expectation value of any operator 
acting on an $N$ dimensional Hilbert space. The expectation value is 
obtained by measuring the polarization of a 
single auxiliary qubit. As an example, we showed how to program 
the evaluation of sums of values of the discrete Wigner function 
over various simple phase space domains. It is important to mention that our 
method is only efficient to determine if the sum of the Wigner 
function in a phase space domain (with up to $o(N)$ points) is greater than 
a fixed, $N$--independent, threshold (since this does not require exponential 
precision). This is a ``quantum decision problem'' whose input data 
(encoded in the system's state $\rho$) is inherently quantum. 
Due to the nature of the input data, this 
problem cannot even be formulated on a classical computer. 
Interest on problems with quantum input data have recently increased, 
partly due to their significance in connection with the potential detection 
of entanglement \cite{Horodecki,Ekert,Horodecki2} as well as their 
relation with tomographic problems like the one described here. 
The extension of some of the above results to 
continuous variables is still under investigation \cite{Horodecki2}.
After completing this work we became aware of the related approach to 
the construction of quantum universal detectors presented in \cite{Dariano2}.

We acknowledge useful discussion with Marcos Saraceno. This work was 
partially supported with grants from Ubacyt, Anpcyt, Conicet 
and Fundaci\'on Antorchas.

%%%%%%%%%%%%%%%%%%%%%%%%%%%%%%%%%%%%%%%%%%%%%%%%%%%%%%%%%%%  
%  
%		Figure captions  
%  
%%%%%%%%%%%%%%%%%%%%%%%%%%%%%%%%%%%%%%%%%%%%%%%%%%%%%%%%%%%  

\end{document}